\documentclass[a4paper]{jpconf}
\usepackage{graphicx,cite}

\usepackage{epsfig,multicol,multirow,cite}
\usepackage{amsmath,subfigure,latexsym,amssymb}

\newcommand{\be}{\begin{equation}}
\newcommand{\ee}{\end{equation}}

\newcommand{\bea}{\begin{eqnarray}}
\newcommand{\eea}{\end{eqnarray}} 

\newcommand{\la}{\langle}
\newcommand{\ra}{\rangle}

\newcommand{\Z}{\mathbb{Z}}
\newcommand{\R}{{\kern+.25em\sf{R}\kern-.78em\sf{I} \kern+.78em\kern-.25em}}
\newcommand{\RR}{{\kern+.25em\sf{R}\kern-.6em\sf{I} \kern+.6em\kern-.25em}}
\newcommand{\N}{{\kern+.25em\sf{N}\kern-.78em\sf{I} \kern+.78em\kern-.25em}}
\newcommand{\C}{{\kern+.25em\sf{C}\kern-.50em\sf{I} \kern+.50em\kern-.25em}}

\newcommand{\vp}{\varphi}

\makeatletter
\@addtoreset{equation}{section}
\makeatother

\begin{document}
\title{Interpretation of topologically restricted measurements
in lattice $\sigma$-models}

\author{Irais Bautista$^{\rm a}$, Wolfgang Bietenholz$^{\rm a}$, 
Urs Gerber$^{\rm a}$,\\
Christoph P.\ Hofmann$^{\rm b}$, H\'{e}ctor Mej\'{\i}a-D\'{\i}az$^{\rm a}$ 
and Lilian Prado$^{\rm a}$}

\address{$^{\rm a}$ Instituto de Ciencias Nucleares,
Universidad Nacional Aut\'{o}noma de M\'{e}xico \\
\ \ \ A.P.\ 70-543, C.P.\ 04510 M\'{e}xico, Distrito Federal, Mexico \\
$^{\rm b}$ Facultad de Ciencias, Universidad de Colima \\
\ \ \ Bernal D\'{\i}az del Castillo 340, Colima C.P.\ 28045, Mexico}

\ead{wolbi@nucleares.unam.mx}

\begin{abstract} 
We consider models with topological sectors,
and difficulties with their Monte Carlo simulation.
In particular we are concerned with the situation where a
simulation has an extremely long auto-correlation time with 
respect to the topological charge. Then reliable numerical 
measurements are possible only within single topological 
sectors. The challenge is to assemble such restricted measurements 
to obtain an approximation for the full-fledged result, 
which corresponds to the correct sampling over the entire set 
of configurations. Under certain conditions this is possible, 
and it provides in addition an estimate for the topological 
susceptibility $\chi_{\rm t}$. Moreover, the evaluation of $\chi_{\rm t}$
might be feasible even from data in just one topological sector,
based on the correlation of the topological charge density.
Here we present numerical test results for these techniques
in the framework of non-linear $\sigma$-models.
\end{abstract}

\section{Topological sectors in quantum physics}

We are going to consider quantum physics in the functional
integral formulation with Euclidean time. For a number
of models, the paths --- or configurations --- to be summed
over split into {\em topological sectors.} The simplest example 
is the quantum rotor, {\it i.e.}\ a quantum mechanical scalar
particle moving freely on a circle with periodic boundary 
conditions in time (say with period $T$).
This corresponds to the 1d $O(2)$ model
with the action, partition function and expectation values
\be
S[\vp ] = \frac{\beta}{2} \int_{0}^{T} \dot 
\vp^{2} \, dt \ , \quad
Z = \int D\vp \ \exp (-S[\vp ]) \ , \quad 
\la {\cal O} \ra = \frac{1}{Z} \int D \vp \ {\cal O}[\vp ] 
\, \exp (-S[\vp ]) \ .
\ee
($\beta$ represents the inverse coupling for the $O(2)$ model,
or the moment of inertia for the rotor.)
The functional integral $\int D \vp$ sums over all closed
paths $\vp (t) \in S^{1}$, $\vp (0) = \vp (T)$. These paths
occur in disjoint subsets, which can be labelled by their
winding number, or {\em topological charge}
\be
Q = \frac{1}{2 \pi} \int_{0}^{T} \dot \vp \ dt \in \Z \ .
\ee
A continuous deformation of a path does not change $Q$, 
hence it remains in the same subset, {\it i.e.}\ in the same 
topological sector.\\

This can be generalised to quantum field theory on a torus, 
namely to $O(N)$ models in $d = N-1$ dimensions, where the 
field is a classical spin $\vec S (x) \in S^{N-1}$. For the
models in this class, the configurations are structured in 
topological sectors. We are going to address the cases $N=2$ and $3$.
Further models with topological sectors are {\it e.g.}\
the 2d $CP(N-1)$ models, 2d $U(1)$ gauge theory, and 4d Yang-Mills
theories. In all these cases, the configurations can only
be deformed continuously within a fixed topological sector.
The functional integral splits into disjoint integrals
over these sectors, which are characterised by a topological
charge $Q \in \Z$.

\section{Lattice regularisation and Monte Carlo simulation}

We proceed to the lattice formulation, which reduces
the (periodic) volume to a set of discrete sites $x$,
say with a cubic structure and lattice spacing $a$. Matter
fields live on these sites, $\Phi_{x}$, while gauge fields
can be given by compact link variables $U_{x,\mu} \in
\{ {\rm gauge~group} \}, \mu = 1 \dots d$. This represents 
a gauge invariant UV regularisation.

On the lattice regularised level, there are {\it a priori} no
topological sectors; all configurations can be continuously 
deformed into one another. Still, something similar exists:
the action may have local minima, and their vicinities are separated
by regions of high action. Hence transitions between these sectors
are statistically suppressed, where 
\be
p [\Phi ]  = \exp ( - S [\Phi ]) / Z
\label{propconf}
\ee
is interpreted as the probability for some configuration $[\Phi ]$. 
In such cases, it is often convenient to introduce a topological
charge $Q$ even on the lattice. In the presence of chiral fermions
(with a lattice Dirac operator that obeys the Ginsparg-Wilson Relation),
this can be done best by the fermion index \cite{Has98}. For 
$(N-1)$-dimensional $O(N)$ models, the optimal formulation maps local 
patches of the lattice configuration onto $S^{N-1}$, and the 
(hyper-)spherical area represents the topological charge density 
(``geometric definition'') \cite{BergLuscher}. 
These definitions both have the virtue that they provide integer
$Q$ values for each configuration (up to a subset of measure zero), 
even on the lattice.

Monte Carlo simulations generate a sequence of configurations 
(a ``Monte Carlo history'') randomly 
with the probability given in eq.\ (\ref{propconf}). 
Summing over this set provides an approximation for expectation
values of observables, such as $n$-point functions. There are still
statistical errors (the set of configurations is
finite), and systematic errors (for instance due to the finite 
lattice spacing $a > 0$),
which can be reduced with increased computational effort. On the other
hand, we stress that this method is fully {\em non-perturbative.}

\section{Topologically frozen numerical measurements}

Most popular algorithms generate configurations by performing
a long sequence of local updates, until a new, (quasi-)independent
configuration emerges, to be used for the numerical measurement.
This includes in particular the Hybrid Monte Carlo (HMC) algorithm 
\cite{HMC}, which is standard in QCD simulations with dynamical quarks.

Regarding theories with topological sectors, however, such local
update algorithms are plagued by the problem that they have a hard time
to change the topological sector. This requires the sequence of update 
steps to pass through a region of low probability. However, in order
to sample the entire space of configuration correctly, the algorithm
is supposed to change $Q$ frequently. As a striking counter-example, we 
mention the QCD study with chiral quarks reported in Ref.\ \cite{JLQCD}, 
where the extensive HMC history was entirely confined to
the sector with $Q=0$. For Wilson-type quarks the
problem is less severe so far, {\it i.e.}\ in the range of lattice
spacings $0.05~{\rm fm} \lesssim a \lesssim 0.15~{\rm fm}$ which
have typically been used up to now. However, once we push for even
finer lattices --- in order to suppress further the systematic lattice
artifacts --- the problem will become more severe: it is welcome for
many reasons that the system is more continuum-like, but this also
makes topological transitions harder. 

Therefore Monte Carlo histories which are trapped in one topological
sector for a very long (computing) time, {\it i.e.}\ over a huge
number of update steps, are a serious issue. The question
arises if we can still perform correct numerical measurements 
of $n$-point functions, or {\it e.g.}\ of the topological susceptibility
\be
\chi_{\rm t} = ( \la Q^{2} \ra - \la Q \ra ^{2}) / V \ .
\ee
As a remedy, the use of open boundary conditions has been advocated
\cite{openbc}; then $Q$ is not integer anymore, so it can vary
continuously. Here we consider attempts to deal with this problem
while keeping the boundary conditions periodic. We test two procedures
for this purpose, with numerical data for the 1d $O(2)$ and the 
2d $O(3)$ model, and we refer to the geometric definition of the
topological charge of a lattice configuration. Its contributions from 
some local patches (nearest neighbour sites in $d=1$, triangles in 
$d=2$) are the lattice topological charge density 
$q$.

\section{Correlation of the topological charge density}

First we consider a method suggested in Ref.\ \cite{AFHO}
to determine $\chi_{\rm t}$ even if only one single topological
sector has been explored in the simulation. We denote the
restricted expectation value within one sector as 
$\la \dots \ra_{Q}$.
 Ref.\ \cite{AFHO} derived an approximate formula for the
correlation of the topological lattice charge density 
at large Euclidean time separation $t$,\footnote{On the
right-hand-side there is another correction term, 
$( \la Q^{4}\ra - 3 \la Q^{2}\ra ^{2}) / (2 V \chi_{\rm t}^{2})$,
for deviations from Gauss-distributed charges $Q$ (``kurtosis''),
but we checked that it is negligible in the examples that we considered.}
\be  \label{japeq}
^{\, \lim}_{t \to \infty} \, \la \ q_{0} \, q_{t} \, \ra_{|Q|} \approx
- \frac{1}{V} \chi_{\rm t} + \frac{Q^{2}}{V^{2}} \ .
\ee
The derivation assumes the charges $Q$ to be Gauss distributed
around zero, $\la Q^{2} \ra = V \chi_{\rm t}$ to be large, and 
$|Q|/\la Q^{2} \ra$ to be small. The explicit meaning of these 
requirements remains to be investigated numerically.\footnote{Actually 
there is already a successful determination of $\chi_{\rm t}$ in 
2-flavour QCD along these lines \cite{Aoki08}, but it used the flavour 
singlet pseudo-scalar density, as suggested in Ref.\ \cite{etaprime}.}
Here we give preliminary results of our ongoing study \cite{prep}. 
For the 1d $O(2)$ model we consider the standard lattice action
(we will call it simply ``standard action'')
and the Manton action \cite{Manton}. For the sum over all
topological sectors, analytical results are given in 
Ref.\ \cite{rot97}, and additional numerical results 
in Ref.\ \cite{rot07}. For the 2d $O(3)$ model
we simulated the standard action (in lattice units)
\be
S [\vec S] = \beta \sum_{x, \mu} (1 - \vec S_{x} \cdot 
\vec S_{x + \hat \mu}) \ ,
\ee
where $\hat \mu$ is a lattice unit vector in $\mu$-direction.\footnote{The 
Manton action replaces the function $\cos \Delta \varphi_{x,\mu}$ 
of the relative angle (in the last term) by 
$\frac{1}{2} (\Delta \varphi_{x,\mu})^{2}$ . 
For the quantum rotor this reduces the lattice artifacts drastically;
this action turns out to be classically perfect \cite{rot97}.}
We used an efficient cluster algorithm \cite{Wolff}, which is not
restricted to local update steps, and which does therefore change
the topological sector frequently. Hence in these tests correctly
distributed configurations are available, and they provide a check
for the indirect methods. The latter are relevant for instance in
gauge theories, where no efficient cluster algorithm is known.

\begin{figure}[hbt!]
\center
\includegraphics[angle=270,width=0.51\linewidth]{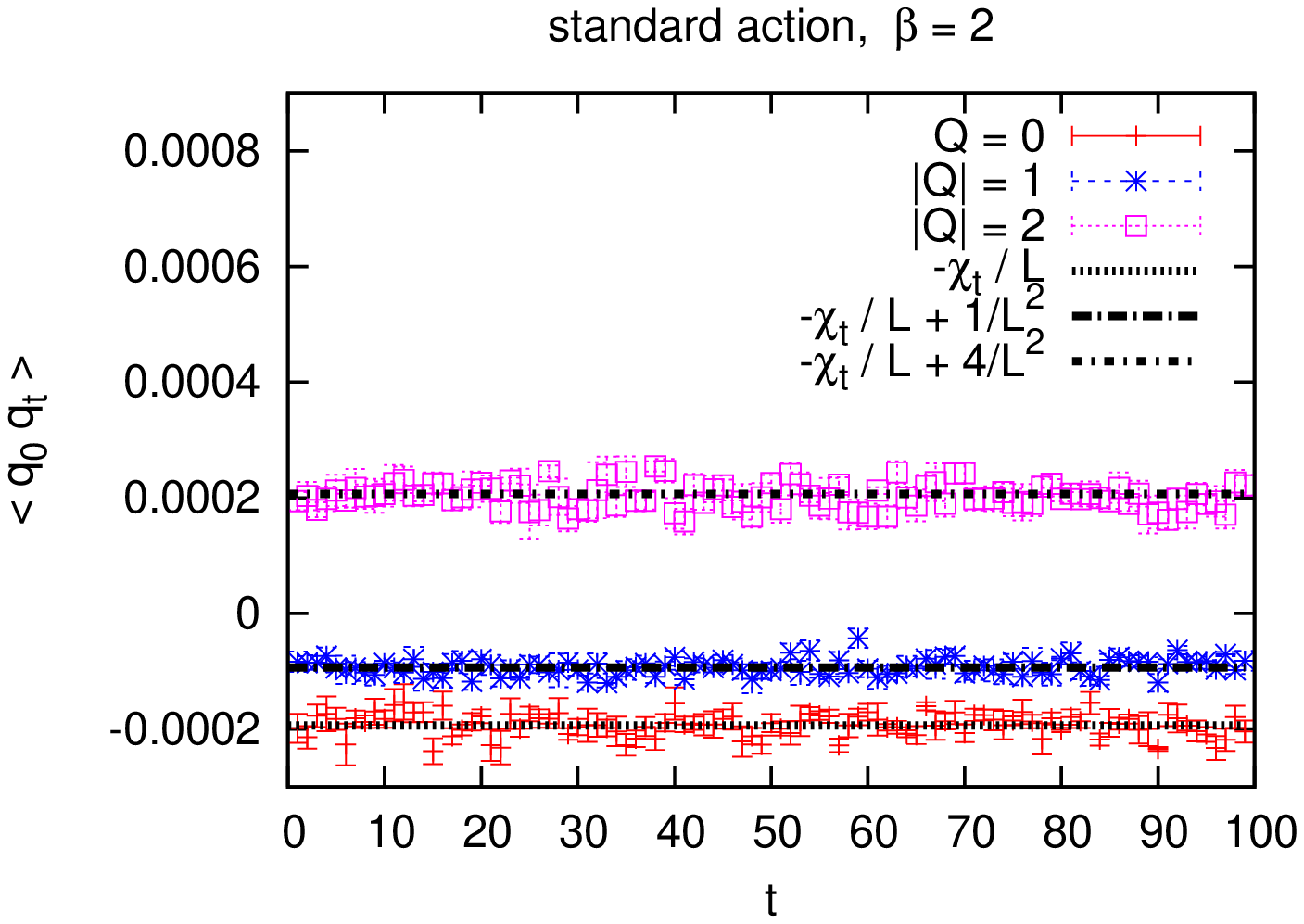}
\hspace{-6mm}
\includegraphics[angle=270,width=0.51\linewidth]{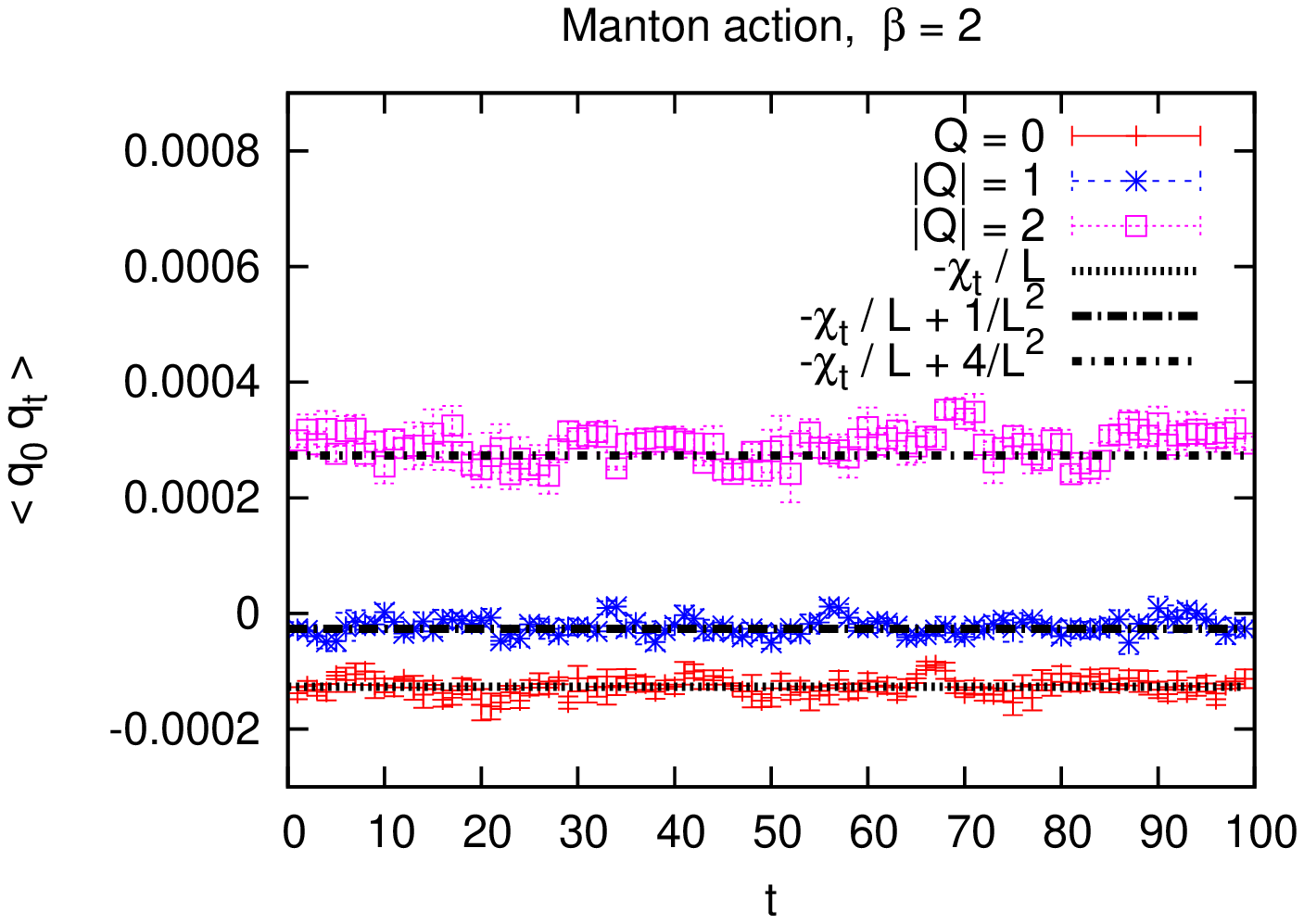}
\caption{The correlation function of the topological charge density
over Euclidean time separation $t$ for the 1d $O(2)$ model on a spin 
chain of length $L=100$ and inverse coupling $\beta = 2$. 
We show results for the standard action (with $\la Q^{2} \ra = 1.94$)
on the left, and for the Manton action (with 
$\la Q^{2} \ra = 1.27$) on the right. The black dotted lines 
correspond to the approximation (\ref{japeq}) for $|Q|=0,1,2$, if we 
insert the analytically known values for $\chi_{\rm t}$ \cite{rot97}.}
\label{corre1d}
\end{figure}

\begin{figure}[hbt!]
\center
\includegraphics[angle=270,width=0.51\linewidth]{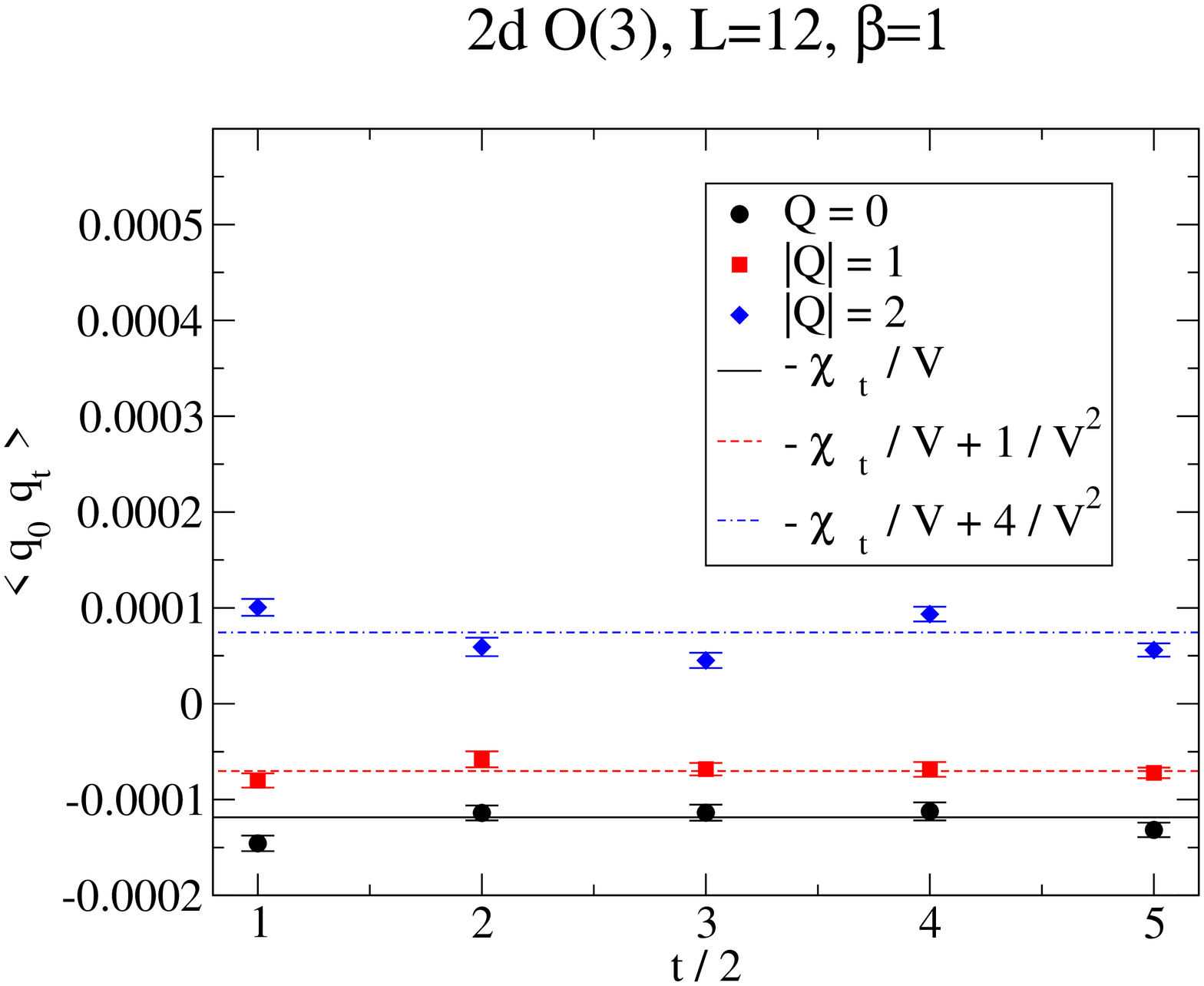}
\hspace{-6mm}
\includegraphics[angle=270,width=0.51\linewidth]{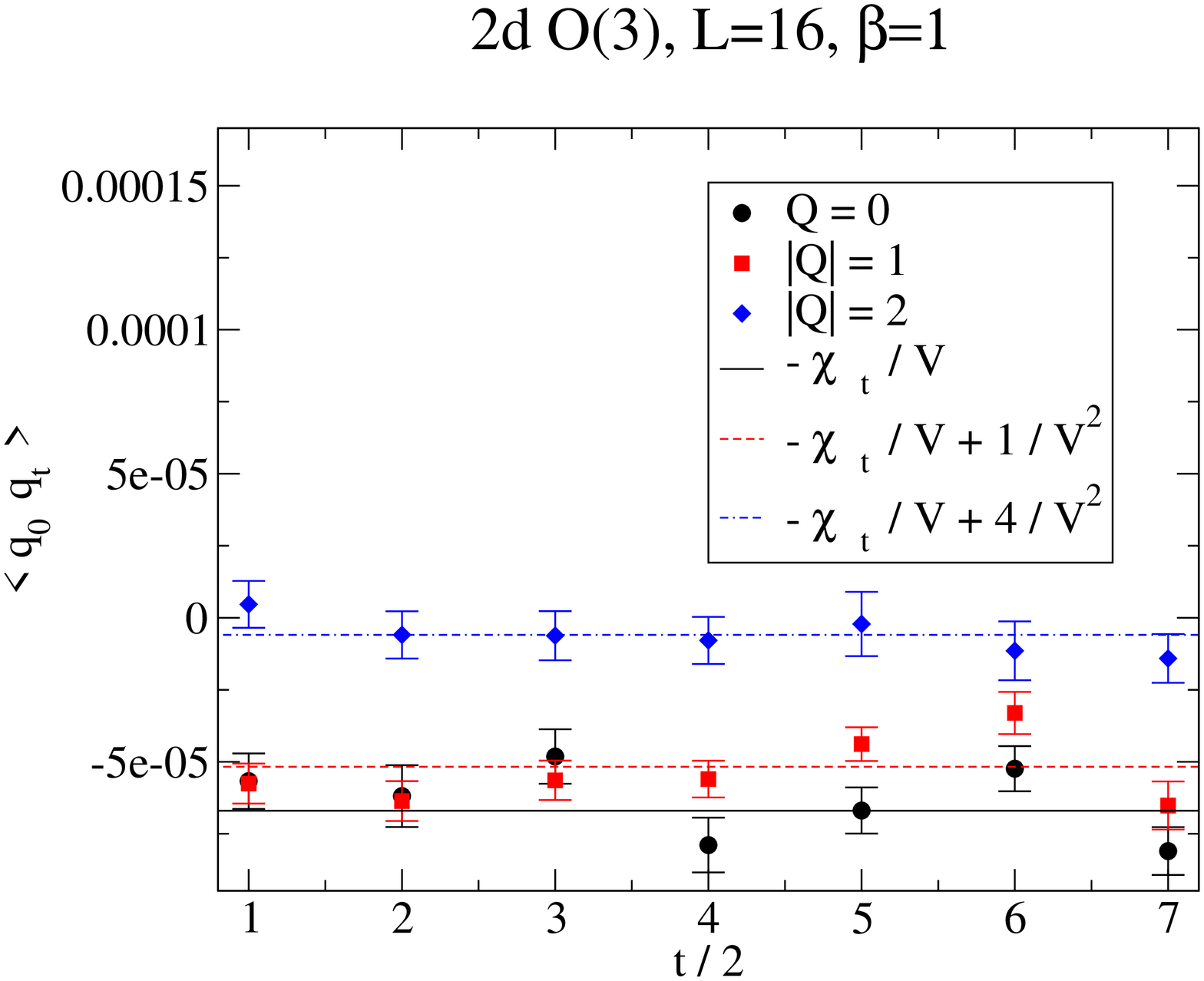}
\caption{The correlation function of the topological charge density
over Euclidean time separation $t$ for the 2d $O(3)$ model at $\beta = 1$. 
We show results for the standard action in volume $V = 12 \times 12$
(with $\la Q^{2} \ra = 2.46$) on the left, and $V = 16 \times 16$
(with $\la Q^{2} \ra = 4.39$) on the right. The horizontal 
lines are the values for $|Q|=0,\, 1$ and $2$ obtained from
the approximation (\ref{japeq}), if we insert the directly measured 
topological susceptibility $\chi_{\rm t}$.}
\label{corre2d}
\end{figure}

Our results for $\la \, q_{0} \, q_{t} \, \ra_{|Q|}$
are shown in Figures \ref{corre1d} and \ref{corre2d}. 
In all cases we observe excellent agreement with the values corresponding 
to the approximation formula (\ref{japeq}). However, in the
field theoretic example of Figure \ref{corre2d} the correlation
length is short ($\xi \simeq 1.3$). To move closer
to the continuum limit, we have to increase $\beta$, and ---
in order to keep $\la Q^{2} \ra$ relatively large --- also the
volume $V$. Consequently the predicted plateaux for
$\la \, q_{0} \, q_{t} \, \ra_{|Q|=0,1,2}$ move rapidly towards zero, and 
it will be difficult to disentangle them in the numerical data from 
each other and from zero. However, this distinction is crucial for
the evaluation of $\chi_{\rm t}$. The question how far this is still 
feasible is currently under investigation. Of course, this issue is 
even more severe in higher dimensions.

\section{Approximate topological summation of observables}

Now we proceed to a more ambitious goal: the computation of some
observable $\la {\cal O} \ra$, if only a few topologically restricted
measurements $\la {\cal O} \ra_{|Q|}$ are available. An approximate
formula for this purpose was first derived in Ref.\ \cite{BCNW},
referring to the pion mass in QCD simulations (for later considerations,
see Refs.\ \cite{AFHO,BHSV,Dromard,Czaban}). It can also be 
applied to other observables ${\cal O}$, and other models with 
topological sectors. The relevant approximation formula reads
\be  \label{sumeq}
\la {\cal O} \ra_{|Q|} \approx \la {\cal O} \ra +
\frac{c}{2 V \chi_{\rm t}} \Big( 1 - \frac{Q^{2}}{V \chi_{\rm t}} \Big) \ ,
\ee
where $c$ is a constant, which (like $\chi_{\rm t}$ and
$\la {\cal O} \ra \, $) stabilises in large volumes.
Together with $\la {\cal O} \ra$ and $\chi_{\rm t} $, there are 
three unknown terms on the right-hand-side.
If we employ numerical results for the left-hand-side, in various
$|Q|$ and $V$, they can in principle be determined by a
fit (this requires results in at least two different volumes).

The assumptions in the derivation are (qualitatively) the same
as for eq.\ (\ref{japeq}), hence we need again a reasonably large
$\la Q^{2} \ra = V \chi_{\rm t}$, and we should only use rather
small charges $|Q|$. Up to now there are only few tests of this 
approximation with numerical data; they were performed in the 
Schwinger model \cite{BHSV,Czaban} and for the quantum rotor 
\cite{Dromard}.

\begin{figure}[hbt!]
\vspace*{-5mm}
\center
\includegraphics[angle=270,width=0.6\linewidth]{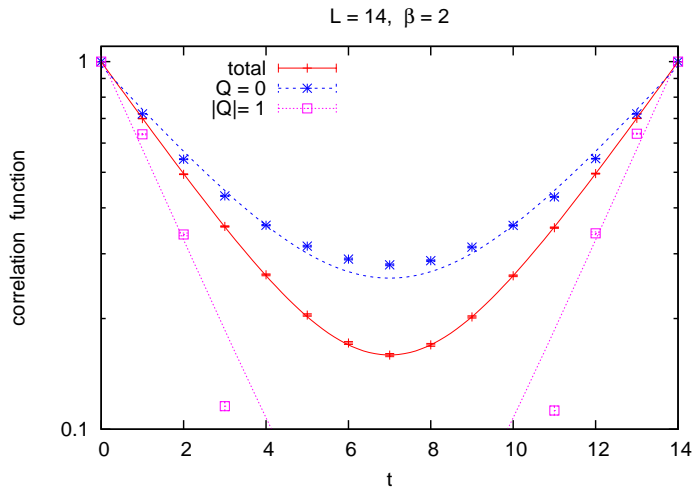}
\caption{The correlation function for the 1d $O(2)$ model with $L=14$,
$\beta =2$ for the sectors $|Q|=0$ and $1$, and in total. Only the
complete correlation follows well a fit to a $\cosh$ function.}
\label{nocosh}
\end{figure}

First we present further results for the latter, {\it i.e.}\ the
1d $O(2)$ model, where we measured the restricted
correlation length $\xi_{|Q|}$ for $|Q|=0,1,2$.
Actually the (connected) correlation function
exhibits a neat exponential decay --- resp.\ a $\cosh$ shape --- 
only if all sectors are included. Topological restriction
deforms this decay somewhat, hence the evaluation of $\xi_{|Q|}$
is less obvious \cite{Dromard,Czaban}.
To illustrate this property, we show in Figure \ref{nocosh} the 
correlation function in a small volume of $L=14$ at $\beta =2$, 
for $|Q|=0$ and $1$, and in total. We see that only the total
(unrestricted) correlation matches a $\cosh$ function very well.

Now we increase the inverse coupling to $\beta =4$, and we
consider larger sizes in the range $L = 150 \dots 400$.
Once we have evaluated $\xi_{|Q|}$ as well as possible,
we use $\xi_{0}$ as an input for the 
summation formula (\ref{sumeq}) to obtain approximations 
$\xi_{\rm 1, approx}$ and $\xi_{\rm 2, approx}$, which are then
compared to the directly measured values. 
\begin{figure}[hbt!]
\center
\includegraphics[angle=270,width=0.51\linewidth]{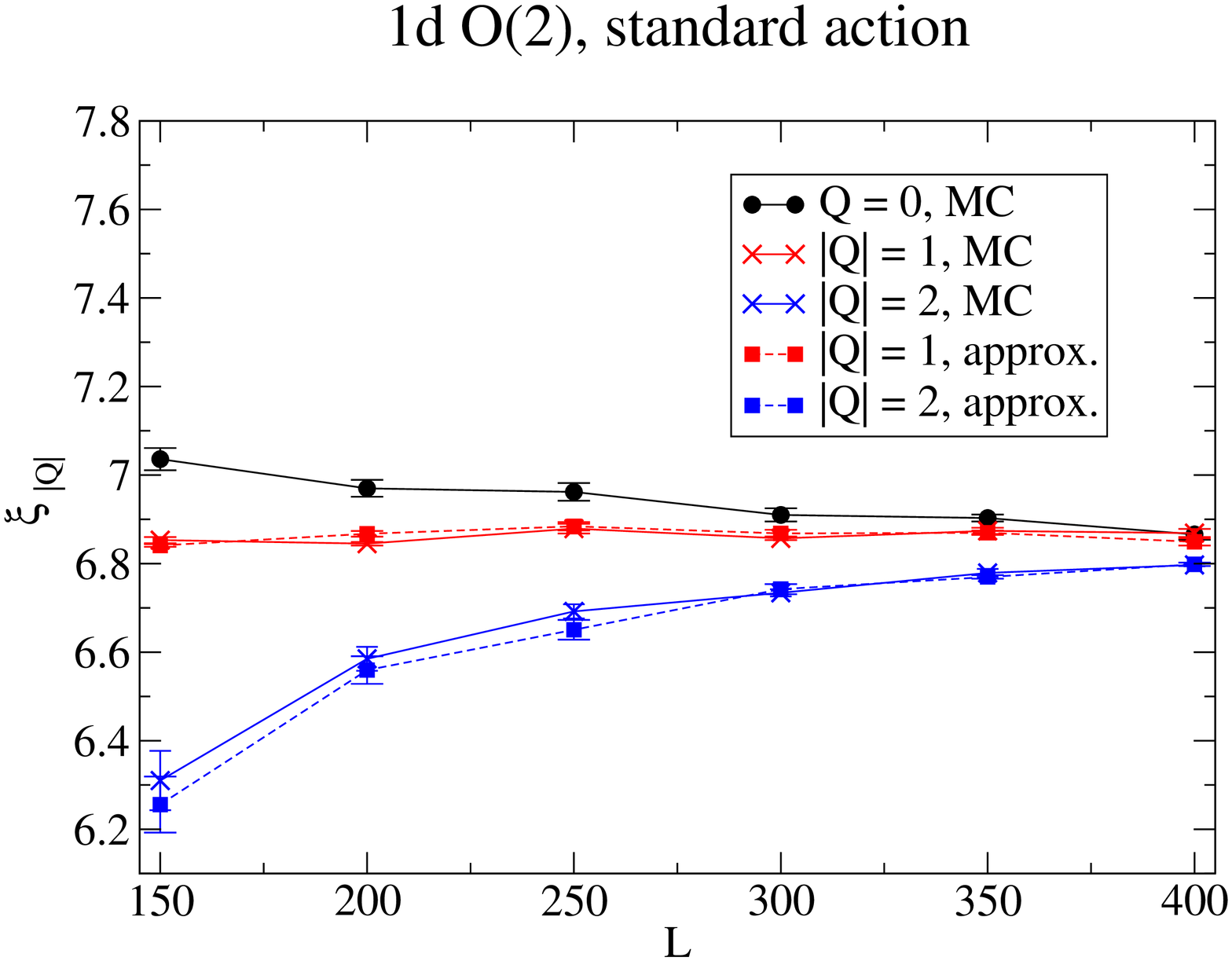}
\hspace{-6mm}
\includegraphics[angle=270,width=0.51\linewidth]{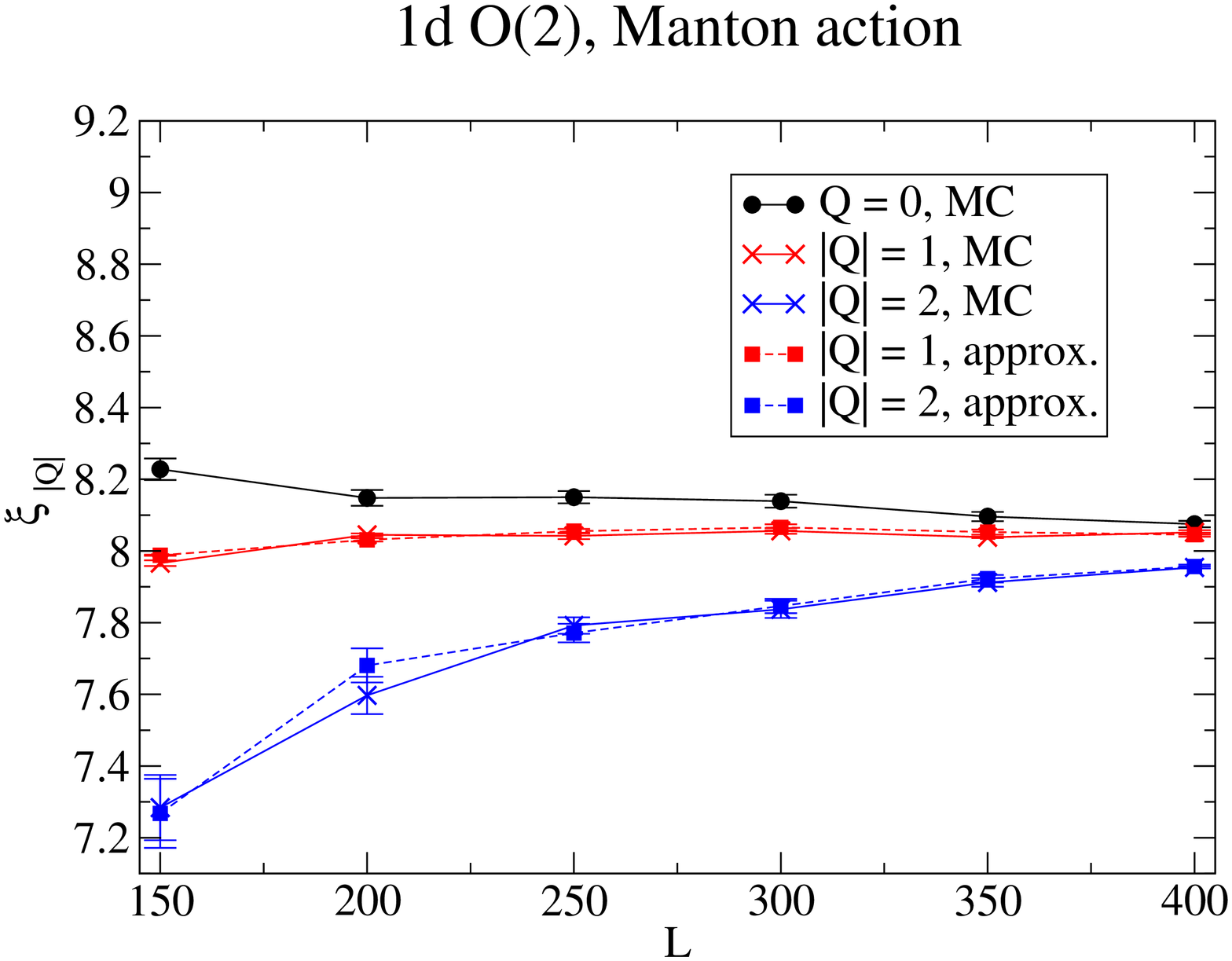}
\caption{The topologically restricted correlation lengths
$\xi_{0}$,  $\xi_{1}$, $\xi_{2}$ measured directly in Monte Carlo
(MC) simulations of the 1d $O(2)$ model at $\beta =4$. 
For $\xi_{1}$ and $\xi_{2}$ these results are 
compared to the values obtained from the approximation formula
(\ref{sumeq}), if $\xi_{0}$ is used as an input. We observe good
agreement in all cases.}
\label{xi1xi2}
\end{figure}
Figure \ref{xi1xi2} shows good agreement over a broad range of $L$ 
both for the standard action and for the Manton action. In particular
for the smaller sizes $L$, this observation is highly non-trivial
since there is a considerable splitting between the lengths $\xi_{|Q|}$.
Of course, in very large volumes all values coincide, 
so this agreement becomes trivial.

Now we address the true correlation length $\xi$, which describes
the decay of the correlation function measured over all topological
sectors. Generally, in very large volumes all restricted values
also coincide with the complete result, as eq.\ (\ref{sumeq}) 
shows, hence also this agreement becomes trivial. 
However, such large volumes are often inaccessible in Monte 
Carlo simulations. Therefore we explore whether in somewhat smaller 
volumes the correction terms on the right-hand-side of eq.\ (\ref{sumeq})
can help to still extract the correct result. The volume should
not be too small, however, otherwise additional corrections, which
are missing in approximation (\ref{sumeq}), become 
significant.\footnote{An extension to higher-order corrections 
is discussed in Ref.\ \cite{Dromard}.}

\begin{figure}[hbt!]
\center
\includegraphics[angle=270,width=0.51\linewidth]{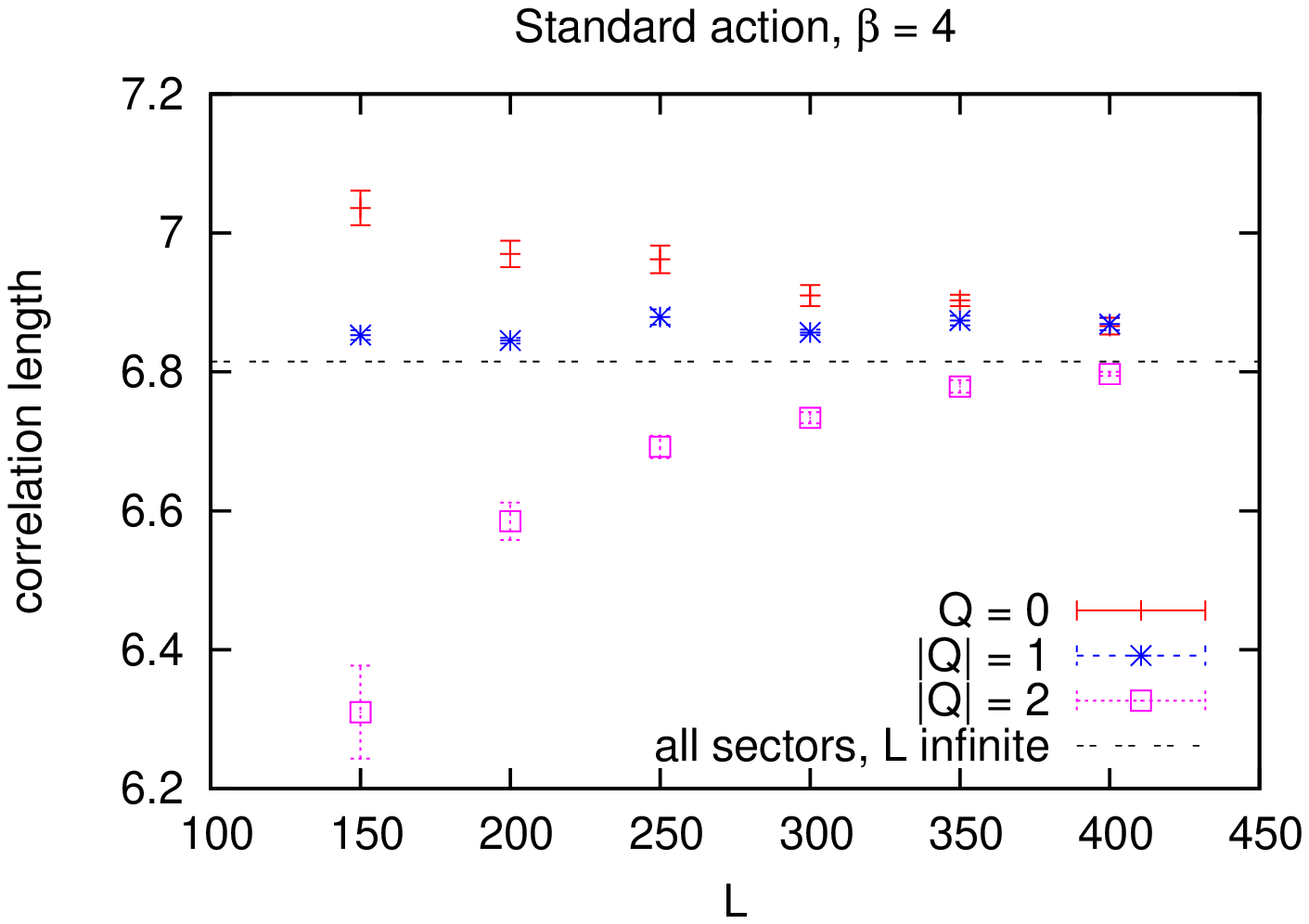}
\hspace{-6mm}
\includegraphics[angle=270,width=0.51\linewidth]{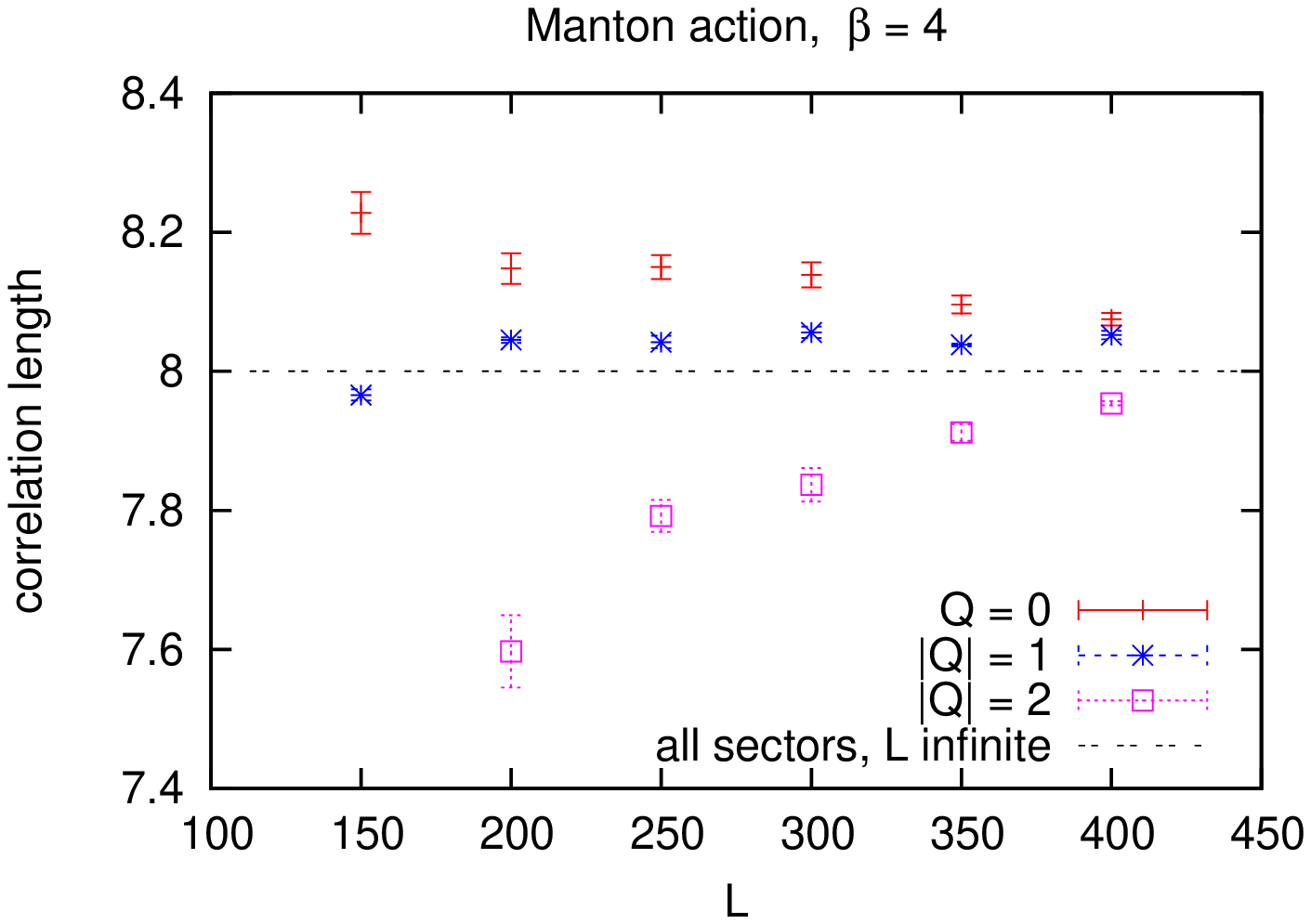}
\caption{The analytic value of the correlation length $\xi$
in the 1d $O(2)$ model (dashed lines),
and numerical measurements for the topologically restricted
correlation lengths $\xi_{0}$,  $\xi_{1}$, $\xi_{2}$ in sizes
$L = 150 \dots 400$. We see a clear splitting both for the
standard action (on the left) and for the Manton action (on the
right). Hence it is non-trivial that these measured values 
provide good estimate for $\xi$, when we insert them into 
approximation (\ref{sumeq}) and perform a least square fit for 
the three unknown terms.
Corresponding results are given in Table \ref{tabxi}.}
\label{xitotal}
\end{figure}

Figure \ref{xitotal} shows the exact correlation length at $\beta =4$,
which is known analytically \cite{rot97} (dashed lines), along with
the numerically measured values for $\xi_{0}$,  $\xi_{1}$, $\xi_{2}$
in sizes $L = 150 \dots 400$. 
\begin{table}[h!]
\centering
\begin{tabular}{|c||c|c|c||c|c|c|c|}
\hline
 & \multicolumn{3}{|c||}{\bf Standard action} &
\multicolumn{3}{|c|}{\bf Manton action} \\ 
\hline
fitting range for $L$ & $ 250 \ - \ 400$ & $ 300 \ - \ 400$ & theory &
$ 250 \ - \ 400$ & $ 300 \ - \ 400$ & theory \\
\hline
\hline
$\xi$ & 6.77(5) & 6.79(2) & 6.815 & 7.95(5) & 7.88(6) & 8.000 \\
\hline
\end{tabular}
\caption{Fitting results for the correlation length $\xi$ in the 
1d $O(2)$ model at $\beta =4$, based on the formula (\ref{sumeq}). We 
insert the numerically measured values for $\xi_{0}$,  $\xi_{1}$, 
$\xi_{2}$ in some range for the size $L$, and determine the optimal 
approximation for $\xi$ by a least-square fit. We arrive
at results which are close to the theoretical value.}
\label{tabxi}
\end{table}
Table \ref{tabxi} gives the results based on eq.\
(\ref{sumeq}), if we insert the numerical and the topologically
restricted values in some range of sizes $L$. This is the procedure 
to handle a situation where the complete result is not known. 
Now the three unknown parameters are strongly over-determined, and
we perform a least-square fit to obtain the estimates in Table
\ref{tabxi}. The fact that there are only small fitting errors
confirms the consistency of this approach. Moreover, the
resulting values are close to the exact values,
which is again non-trivial regarding the significant splitting
of $\xi_{0}$,  $\xi_{1}$ and $\xi_{2}$ (see Figure \ref{xitotal}).\\

\begin{figure}[hbt!]
\vspace*{-5mm}
\center
\includegraphics[angle=270,width=0.6\linewidth]{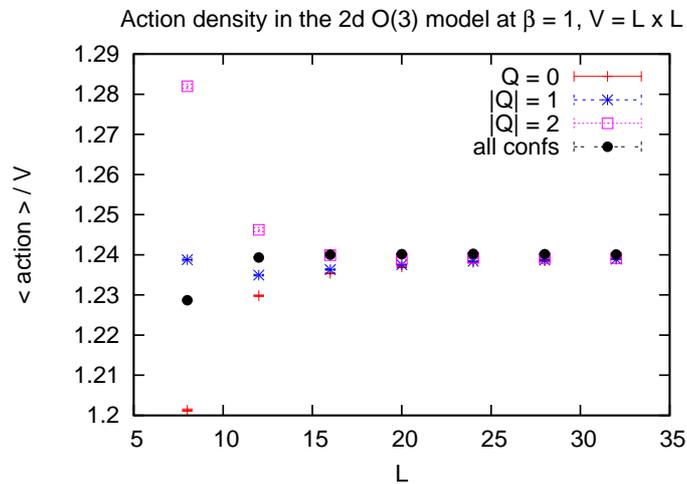}
\caption{Numerical results for the action density of the 2d $O(3)$ 
model with the standard action at $\beta =1$ in various $L \times L$ 
volumes. We show the complete result, as well as the values
restricted to $|Q| = 0,\ 1$ or $2$.}
\label{actdens}
\end{figure}

Finally we add an application of formula 
(\ref{sumeq}) to the 2d $O(3)$ model. Here our observable
is the action density, {\it i.e.}\ the mean value of the action
divided by the volume, $\la S \ra /V$. Figure \ref{actdens}
shows results for the sectors $|Q| = 0,\ 1,\ 2$ and for all 
configurations, in $L \times L $ volumes with 
$L = 8 \dots 32$. We see that they all
converge to the same value for increasing size $L$.

If we were not able to measure the result over all configurations
reliably --- {\it i.e.}\ if we only had Monte Carlo histories that
hardly ever change the topological sector --- we could now use
the restricted measurements in some range of $L$, insert them
into formula (\ref{sumeq}) and estimate $\la S \ra /V$, as well
as $\chi_{\rm t}$, with a least-square fit. Such results are
displayed in Table \ref{2dO3tab}. 
\begin{table}[h!]
\centering
\begin{tabular}{|c||c|c|c||c|}
\hline
& & & &  directly measured \\
fitting range for $L$ & $ 16 \ - \ 24$ & $ 16 \ - \ 28$ & 
$ 16 \ - \ 32$ & in all sectors at $L=32$ \\
\hline
\hline
$\la S \ra / V $ & 1.24038(12) & 1.24027(8) & 1.24015(5) & 1.24008(5) \\
\hline
$\chi_{\rm t}$ & 0.0173(6) & 0.0169(5) & 0.0164(5) & 0.01721(4) \\
\hline
\end{tabular}
\caption{Estimates for the action density $\la S \ra /V$ and for 
the topological susceptibility  $\chi_{\rm t}$ in the 2d $O(3)$ model
at $\beta = 1$, based on topologically restricted results 
in some range of $L$. We find good agreement with the directly 
measured values, without topological restriction (last column).}
\label{2dO3tab}
\end{table}
We see that even the use of modest volumes provides estimates, 
which agree well with the results of a direct complete measurement 
(this is feasible in this case thanks to the cluster algorithm).

\section{Summary}

Most Monte Carlo simulations of lattice field theory 
still have to be performed with local update algorithms.
For theories with topological sectors, this often implies
that the Monte Carlo history is confined to one sector over
an enormous number of update steps --- a problem, which is
getting worse for decreasing lattice spacing. This concerns
in particular QCD simulations with dynamical quarks. 

In this case, it may happen that observables can be measured
well only within a fixed sector, {\it i.e.}\ one is limited
to topologically restricted expectation values $\la {\cal O}\ra_{Q}$. 
On the conceptual side, this rises algorithmic questions 
related to ergodicity. If we trust the restricted 
measurements $\la {\cal O}\ra_{Q}$, the next issue is to 
extract physical information from them. 

An approximate formula derived in Ref.\ \cite{BCNW} allows in 
principle to obtain an estimate for the physical value 
$\la {\cal O}\ra$, based on restricted results in various
topological sectors and in various volumes. Moreover, if the 
Monte Carlo history rarely changes $Q$, a direct measurement of 
the topological susceptibility $\chi_{\rm t}$ is hardly possible. 
A fit to the formula (\ref{sumeq}) of Ref.\ \cite{BCNW} 
provides in addition an estimate for $\chi_{\rm t}$.
That specific quantity can also be estimated from the correlation
of the topological charge density, even within just one sector, based 
on another approximation formula put forward in Ref. \cite{AFHO},
eq.\ (\ref{japeq}).

Here we have tested both formulae in non-linear $\sigma$-models
with topological sectors. In this case, we have the full-fledged
result (involving all topological sectors) 
available for comparison, either analytically or numerically,
hence we could check how far the approximations work. In Section
4 we tested eq.\ (\ref{japeq}) to determine $\chi_{\rm t}$, and
in Section 5 eq.\ (\ref{sumeq}) for the determination of full
expectation values $\la {\cal O}\ra$ and $\chi_{\rm t}$.

Our results show that these formulae do work well in a
suitable regime, where the assumptions for their derivations are 
decently satisfied. In particular, we need $\la Q^{2}\ra \gtrsim 1.5$,
and we should only involve sectors with $|Q| \leq 2$.
These properties will be further explored in $\sigma$-models
and in Yang-Mills gauge theory \cite{prep}, in order to probe 
the prospects for applications in full QCD.

\ack We thank Christopher Czaban, Arthur Dromard, Ivan Hip and 
Marc Wagner for helpful communication and collaboration.
This work was supported by the Mexican {\it Consejo Nacional de Ciencia 
y Tecnolog\'{\i}a} (CONACyT) through project 155905/10 ``F\'{\i}sica 
de Part\'{\i}culas por medio de Simulaciones Num\'{e}ricas'', as well 
as DGAPA-UNAM. The simulations were performed on the cluster of the 
Instituto de Ciencias Nucleares, UNAM.\\


\begin{thebibliography}{10}

\bibitem{Has98} Hasenfratz P, Laliena V and Niedermayer F
1998 {\it Phys.\ Lett.}\ B {\bf 427} 125

\bibitem{BergLuscher} Berg B and L\"{u}scher M 1981
{\em Nucl.\ Phys.}\ B {\bf 190} 412

\bibitem{HMC} Duane S, Kennedy A D, Pendleton B J
and Roweth D 1987 
{\em Phys.\ Lett.}\ {\bf 195} B 216

\bibitem{JLQCD} Fukaya H {\it et al} 2007
{\it Phys.\ Rev.\ Lett.}\ {\bf 98} 172001

\bibitem{openbc} L\"{u}scher M 2010
{\it JHEP} {\bf 1008} 071

\bibitem{AFHO} Aoki S, Fukaya H, Hashimoto S and Onogi T 2007
{\it Phys.\ Rev.}\ D {\bf 76} 054508


\bibitem{Aoki08} Aoki S {\it et al} 2008
{\it Phys.\ Lett.}\ B {\bf 665} 294

\bibitem{etaprime} Fukaya H and Onogi T 2004
{\it Phys.\ Rev.}\ D {\bf 70} 054508

\bibitem{prep} Bautista I, Bietenholz W, Czaban C, Dromard A, 
Gerber U, Hofmann C P, Mej\'{\i}a-D\'{\i}az H, Prado L and 
Wagner M, in preparation

\bibitem{Manton} Manton N 1980 {\em Phys.\ Lett.}\ B {\bf 96} 328

\bibitem{rot97} Bietenholz W, Brower R, Chandrasekharan S
and Wiese U-J 1997
{\it Phys.\ Lett.}\ B {\bf 407} 283

\bibitem{rot07} Boyer T, Bietenholz W and Wuilloud J 2007
{\it Int.\ J.\ Mod.\ Phys.}\ C {\bf 18} 1497

\bibitem{Wolff} Wolff U 1989 {\em Phys.\ Rev.\ Lett.}\ {\bf 62} 361

\bibitem{BCNW} Brower R, Chandrasekharan S, Negele J W
and Wiese U-J 2003 {\it Phys.\ Lett.}\ B {\bf 560} 64

\bibitem{BHSV} Bietenholz W and Hip I 2008
{\em PoS LATTICE2008} 079;
2012 {\em J.\ Phys.\ Conf.\ Ser.}\ {\bf 378} 012041 \\
Bietenholz W, Hip I, Shcheredin S and Volkholz J 2012
{\em Eur.\ Phys.\ J.}\ C {\bf 72} 1938

\bibitem{Dromard} Dromard A and Wagner M,
arXiv:1309.2483 [hep-lat]

\bibitem{Czaban} Czaban C and Wagner M,
arXiv:1310.5258 [hep-lat]

\end{thebibliography}
\end{document}